\documentclass[]{spie}  

\usepackage{amsmath,amsfonts,amssymb}
\usepackage{graphicx}
\usepackage{subcaption}
\usepackage[colorlinks=true, allcolors=blue]{hyperref}
\usepackage{svg}
\usepackage{siunitx}

\usepackage{booktabs}   
\usepackage{tabularx}  
\usepackage{array}     
\newcolumntype{L}{>{\raggedright\arraybackslash}X}

\title{Performance Analysis of the Asgard/NOTT Nulling Interferometer: Optimizing Observing Modes for High-contrast Detection}

\author[a]{J. P. Scott}
\author[a, b]{S. Ertel}
\author[a]{T. A. Stuber}
\author[c]{D. Defrère}
\author[c]{R. Laugier}
\author[d]{J. C. Augereau}   
\author[c]{P. Chingaipe}     
\author[e]{G. Garreau}       
\author[f]{X. Haubois}       
\author[g]{M. Ireland}       
\author[h]{F. Kirchschlager} 
\author[i]{S. Kraus}         
\author[j]{F. Martinache}    
\author[j]{M-A. Martinod}    
\author[c]{K. Missiaen}      
\author[j]{S. Morel}         
\author[c]{J. Morren}        
\author[k]{K. Ollmann}       
\author[d]{P. Priolet}       
\author[c]{G. Raskin}        
\author[c]{M. Salman}        
\author[f]{N. Schuhler}      
\author[c]{W. Verstraeten}   
\author[k]{S. Wolf}          

\affil[a]{Department of Astronomy and Steward Observatory, The University of Arizona, 933 North Cherry Avenue, Tucson, AZ 85721, USA}
\affil[b]{Large Binocular Telescope Observatory, The University of Arizona, 933 North Cherry Avenue, Tucson, AZ 85721, USA}
\affil[c]{Institute of Astronomy, KU Leuven, Celestijnenlaan 200D, 3001 Leuven, Belgium}
\affil[d]{Université Grenoble Alpes, CNRS, IPAG, 38000 Grenoble, France}
\affil[e]{ETH Zürich, Institute for Particle Physics \& Astrophysics, Wolfgang-Pauli-Str. 27, 8093 Zürich, Switzerland}
\affil[f]{European Southern Observatory, Alonso de Cordova 3107 Vitacura, 19001 Santiago, Chile}
\affil[g]{Research School of Astronomy \& Astrophysics, Australian National University, Canberra, ACT 2611, Australia}
\affil[h]{Sterrenkundig Observatorium, Ghent University, Krijgslaan 281-S9, B-9000 Gent, Belgium}
\affil[i]{Department of Physics and Astronomy, University of Exeter, Stocker Road, Exeter EX4 4QL, UK}
\affil[j]{Université Côte d'Azur, Observatoire de la Côte d'Azur, CNRS, Laboratoire Lagrange, France}
\affil[k]{Institut of Theoretical Physics and Astrophysics, University of Kiel, Leibnizstr. 15, 24118 Kiel, Germany}

\authorinfo{Further author information:\\e-mail: jscott25@arizona.edu}

\begin{document} 
\maketitle

\begin{abstract}
We evaluate the performance of three beam-combination schemes, single-Bracewell, asymmetric dual-Bracewell, and symmetric dual-Bracewell, for the forthcoming Asgard/NOTT nulling interferometer at the Very Large Telescope Interferometer. Utilizing the SCIFYsim end-to-end simulator, we assess the instrument's performance by deriving the precision of calibrated null measurements as a function of stellar magnitude and simulating observations of varying hot exozodiacal dust (HEZD) distributions and a hot Jupiter. The study reveals distinct trade-offs for each observing mode. The single-Bracewell mode provides high throughput and preserves spatial information but suffers from poor error suppression. The asymmetric dual-Bracewell mode offers the strongest error suppression for detecting point-like sources, but it inherently suppresses symmetric astrophysical signals such as expected from HEZD. The symmetric dual-Bracewell mode provides a middle ground with modest error suppression while being sensitive to symmetric emission. We conclude that utilizing a combination of all three observing modes provides a robust strategy for detecting HEZD, constraining the structure of its distribution, and identifying false positives from stellar companions.
\end{abstract}


\section{INTRODUCTION}
\label{sec:intro}

Hot exozodiacal dust (HEZD)\cite{kral:2017, ertel:2025} is the presumed origin of a $\sim$1\% near-infrared (nIR) excess emission around $\sim$20\% of main-sequence stars detected with optical long-baseline interferometry\cite{absil:2013, nunez:2017, ertel:2014, absil:2021}. Apart from being hot dust ($\sim$\num{1000} – \qty{2000}{\kelvin}), composed of dominantly submicrometer sized, non-silicate grains, and located close to the host star ($\lesssim$\qty{1}{au})\cite{absil:2006, akeson:2009, kirchschlager:2017, stuber:2023b}, this phenomenon remains poorly understood. The dust causes extended circumstellar emission, which may contain a significant visible light component from light scattered off the small dust grains\cite{kirchschlager:2017, ertel:2025}.  If not properly accounted for, this excess emission could result in significant coronagraphic leakage preventing the detection of exo-Earths with high-contrast imaging or interferometry\cite{ertel:2025}. The dust might also pass through the host star's habitable zone (HZ) when being delivered to or removed from the stellar vicinity\cite{pearce:2020, pearce:2022}, causing confusion and excess noise that can prevent the characterization of exo-Earth atmospheres. Thus, HEZD must be better understood to mitigate risks to future exo-Earth missions such as NASA’s Habitable Worlds Observatory (HWO) and the European Large Interferometer For Exoplanets (LIFE)\cite{quanz:2022} concept.

The NOTT (Nulling Observations for exoplaneTs and dusT) instrument\cite{defrere:2018a, defrere:2018b, defrere:2022, defrere:2024, laugier:2023, garreau:2022, garreau:2024a, garreau:2024b, sanny:2026}, part of the Asgard instrument suite,\cite{martinod:2023} is a forthcoming nulling interferometric\cite{bracewell:1978} beam combiner for the Very Large Telescope Interferometer (VLTI)\cite{haubois:2020} that is designed for detecting young giant planets orbiting near the snowline of nearby main-sequence stars\cite{dandumont:2022} and HEZD. While the sensitivity of NOTT's primary beam combination scheme, a dual-Bracewell nuller, has been studied extensively before,\cite{laugier:2023} this mode is specifically sensitive to asymmetric astrophysical targets such as a planet next to its host star. Consequently, this mode is not expected to be ideal for detecting the presumably mostly symmetric circumstellar emission of HEZD. At the same time, even with a naive approach of combining telescopes pair-wise and assuming a nulling performance that is not exceeding existing nulling observations\cite{ertel:2018a, ertel:2020a}, NOTT is expected to be $\sim$50 times more sensitive to HEZD than past surveys.\cite{absil:2013,ertel:2014,absil:2021} This results from a tenfold improvement in instrumental contrast performance compared to constructive optical long-baseline interferometry combined with a fivefold more favorable dust-to-star flux ratio in the $L$~band\cite{kirchschlager:2020} compared to the $H$/$K$~bands\cite{defrere:2022}. At this sensitivity, a NOTT survey for HEZD can produce a HEZD luminosity function, repeated observations can constrain the variability of the excess emission, and a combination with past and quasi-simultaneous observations from other instruments such as PIONIER\cite{ertel:2014} and MATISSE\cite{kirchschlager:2020,ollmann:2025, priolet:inpress} can produce highly-complementary data in terms of wavelength and u-v-plane coverage, albeit less precise, for a detailed characterization of the dust.
Furthermore, by delivering rich spatial information paired with high precision, NOTT will be able to reliably differentiate between HEZD and stellar companions (phenomena that can even coexist\cite{stuber:2026a}), an issue that plagued long-baseline interferometric observations due to flawed methods to reject stellar companions\cite{tsishchankava:2026} and the typical requirement of joining all data to yield a detection.

In this paper, we study the performance of alternative beam-combination schemes for NOTT and compare them to the previously studied dual-Bracewell mode. We focus our efforts on observations with VLTI's Auxiliary Telescopes (ATs), as we found that using the Unit Telescopes (UTs) is typically not required for HEZD studies. The instrument performance is assessed using SCIFYsim,\footnote{https://github.com/rlaugier/SCIFYsim}$^,$\cite{laugier:2023} a comprehensive end-to-end simulator for NOTT, which we have further extended for our study. For each beam combination scheme, we derive the precision of the calibrated astrophysical null measurements as a function of stellar magnitude and derive a limiting magnitude. We then simulate observations of differing dust distributions to determine the respective instrument response. For completeness and illustration, we also simulate an observation of a hot Jupiter. Identifying the ideal beam combination scheme requires comparing the instrument response amplitude (signal) to the precision of the measured null depth (noise) to predict the resulting signal-to-noise ratio (SNR) of an observation. Further, the information content of an observation needs to be considered. Combining all four telescopes to perform a single precise null measurement may yield less information than combining telescopes pair-wise for two simultaneous null measurements.

The paper is organized as follows. Section~\ref{sec:inst} introduces the NOTT instrument, detailing its multistage photonic beam combiner and the specific observing modes evaluated in this study: single-Bracewell, asymmetric dual-Bracewell, and symmetric dual-Bracewell. Section~\ref{sec:performance} presents the simulated performance of these modes using the SCIFYsim software, exploring both the null-depth precision in the presence of realistic noise sources and simulated target null-depth curves for various astrophysical scenarios, including dust disks and a hot Jupiter. Finally, Section~\ref{sec:conclusion} provides concluding remarks, summarizing the performance trade-offs of each observing mode and highlighting the benefits of combining their data to confidently detect and characterize HEZD.

\section{THE NULLING OBSERVATIONS FOR EXOPLANETS AND DUST INSTRUMENT}
\label{sec:inst}

In this Section we discuss the features of NOTT that we specifically exploit for the present study.  For detailed information on the NOTT instrument we refer to the pertinent references\cite{defrere:2018a, defrere:2018b, defrere:2022, defrere:2024, laugier:2023, garreau:2022, garreau:2024a, garreau:2024b, sanny:2026}.

\subsection{NOTT Beam Combiner}
\label{subsec:combiner}

\begin{figure}
    \centering
    \includegraphics[width=0.5\linewidth]{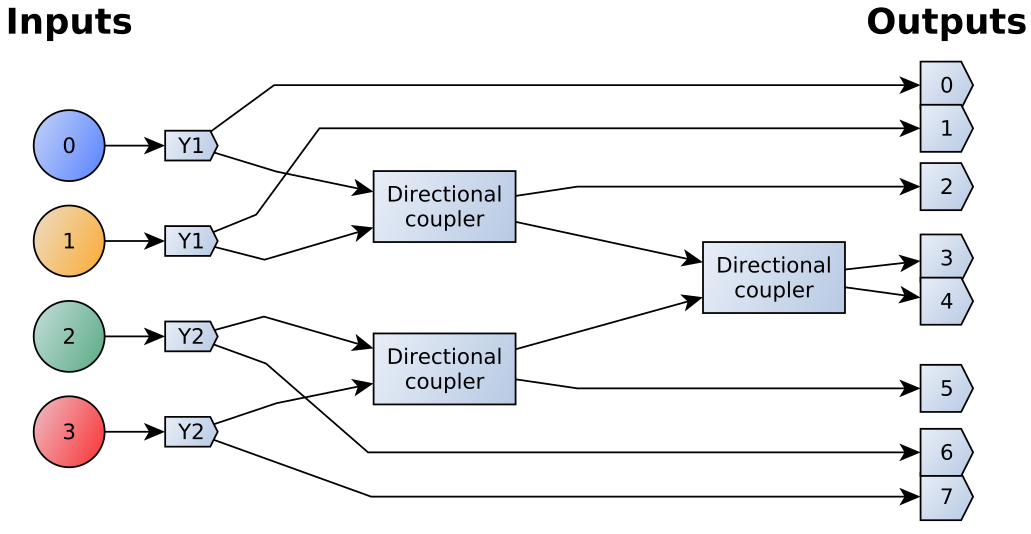}
    \caption{Diagram of the NOTT multistage beam combiner chip\cite{laugier:2023}. Four input beams from the ATs or UTs are combined to produce 4 photometric and 4 interferometric outputs. By adjusting the phases of the input beams, three observing modes can be configured. }
    \label{fig:combiner}
\end{figure}

NOTT contains a multistage photonic beam combiner chip\cite{sanny:2026,garreau:2026} (see Fig.~\ref{fig:combiner}) that accepts four input beams from the VLTI ATs or UTs and produces eight outputs. Input beams are initially split into photometric and interferometric channels. The photometric channels are directed to outputs 0, 1, 6, and 7. The interferometric channels of input beams 0 and 1 propagate to a first-stage directional coupler, which outputs combinations of the input beams. These two outputs feature a relative phase shift of $\pi$\,rad, nominally tuned to the central science wavelength of $\lambda \qty{3.75}{\um}$. A similar first-stage combiner processes input beams 2 and 3. One channel from each of the first-stage directional couplers is directed to outputs 2 and 5, providing pairwise combined interferometric signals. The remaining channels from the first-stage combiners are sent to a second-stage directional coupler. This final coupler produces outputs 3 and 4, again with a relative phase shift of $\pi$\,rad between the inputs, yielding two interferometric outputs that contain information from all six simultaneous baselines. Beam combiner outputs are subsequently spectrally dispersed and imaged on a HAWAII-2RG infrared detector \cite{Hall:2011}.

The final phase combination of the beams at the outputs of the combiner can be tuned by manipulating the relative phase of the input beams by modulating their optical path lengths before injection into the combiner. We have studied three configurations (beam combination schemes) which we identify as distinct observing modes: single-Bracewell mode, asymmetric dual-Bracewell mode, and symmetric dual-Bracewell mode.

\begin{figure}[htbp]
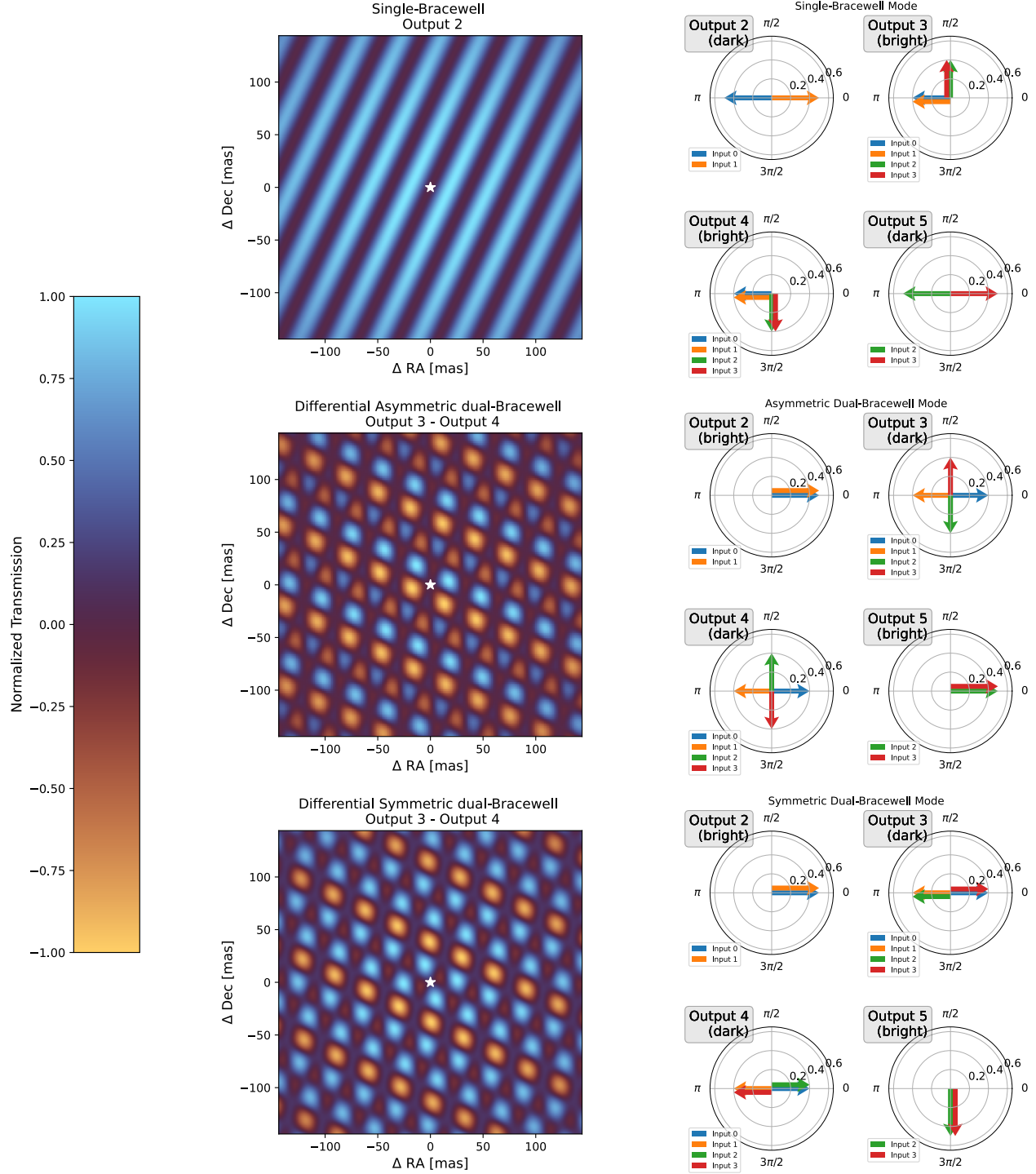

    \centering
    
    \begin{minipage}[b]{0.15\textwidth}
        \centering
        \begin{subfigure}[b]{\textwidth}
            \centering
            \includesvg[height=0.5\textheight]{figures/tran_colorbar.svg}
        \end{subfigure}
        \vspace{3cm}
    \end{minipage}
    \hfill
    \begin{minipage}[b]{0.80\textwidth}
        \centering
        
        \begin{subfigure}[b]{0.46\textwidth}
            \centering
            \includesvg[width=\textwidth, height=\textwidth]{figures/bracewell_mode_diffmap.svg}
        \end{subfigure}
        \hfill
        \begin{subfigure}[b]{0.46\textwidth}
            \centering
            \includesvg[width=\textwidth, height=\textwidth]{figures/bracewell_mode_matrix.svg}
        \end{subfigure}
        
        \vspace{0.1cm} 
        
        \begin{subfigure}[b]{0.46\textwidth}
            \centering
            \includesvg[width=\textwidth, height=\textwidth]{figures/asym_mode_diffmap.svg}
        \end{subfigure}
        \hfill
        \begin{subfigure}[b]{0.46\textwidth}
            \centering
            \includesvg[width=\textwidth, height=\textwidth]{figures/asym_mode_matrix.svg}
        \end{subfigure}
        
        \vspace{0.1cm} 
        
        \begin{subfigure}[b]{0.46\textwidth}
            \centering
            \includesvg[width=\textwidth, height=\textwidth]{figures/sym_mode_diffmap.svg}
        \end{subfigure}
        \hfill
        \begin{subfigure}[b]{0.46\textwidth}
            \centering
            \includesvg[width=\textwidth, height=\textwidth]{figures/sym_mode_matrix.svg}
        \end{subfigure}
    \end{minipage}

    \caption{Transmission maps (left) and output phase diagrams (right) for the three observing modes with the AT small configuration. The phase diagrams display in polar coordinates the amplitude (radial coordinate) and phase (azimuthal coordinate) of each input beam at the individual outputs.  Single-Bracewell mode (top) produces two independent Bracewell nulls at outputs 2 and 5 (only one transmission map is shown). Asymmetric dual-Bracewell mode (middle) produces conjugate nulls that are subtracted. Symmetric dual-Bracewell mode (bottom) produces nulls that are subtracted. The exact phase configurations of the modes on-sky may vary from the examples presented here, however, the examples capture the relevant properties of the null/dark outputs.}
    \label{fig:obsmodes}
\end{figure}

\subsection{Observing Modes}
\label{subsec:obsmodes}

The observing modes can be represented by a complex combiner matrix $\mathbf{M}$ such that
\begin{equation}
    \label{eq:combiner}
    \mathbf{x} = \mathbf{M} \cdot \mathbf{z} \;,
\end{equation}
where $\mathbf{z}$ is a vector representing the electric fields input to the combiner and $\mathbf{x}$ is the output electric field vector. For NOTT, $\mathbf{M}$ is a matrix with eight rows (index $m$) and four columns (index $n$). Each row describes how the input beams are combined to produce an output electric field with the signal on the detector proportional to $\mathbf{I}_m \propto |\mathbf{x}_m|^2$.
The terms of $\mathbf{M}$ have the form 
\begin{equation}
    \label{eq:matrix_terms}
    a_{m,n}(\lambda, \alpha, \beta)\,e^{j\phi_{m,n}(\lambda, \alpha, \beta)} \;,
\end{equation}
where the amplitude $a_{m,n}(\lambda, \alpha, \beta)$ and phase $\phi_{m,n}(\lambda, \alpha, \beta)$ are wavelength dependent and influenced by the optical properties of the combiner chip and VLTI, the relative geometry of the collector array with respect the the sky coordinates $(\alpha, \beta)$, and error terms from atmospheric fluctuations. We use $j$ as the imaginary unit for complex number representation.

In discussing the different observing modes, we are mostly interested in the rows of the combiner matrix associated with the interferometric outputs (i.e., $2 \leq m \leq 5$). The photometric outputs in each mode produce equivalent signals on the detector despite optical path (thus phase) variations in the different modes.

Each output of the beam combiner can be associated with an on-sky transmission map by taking the modulus squared of its associated row in the combiner matrix,
\begin{equation}
    \label{eq:transmission}
    T_m(\lambda, \alpha, \beta) = |\mathbf{M}_m|^2 \;.
\end{equation}
The signal on the detector can then be estimated by integrating the product of the transmission map with the on-sky intensity distribution $J(\lambda, \alpha, \beta)$,
\begin{equation}
    \label{eq:det_signal}
    I_m(\lambda) \propto \sum_{\alpha, \beta} T_m(\lambda, \alpha, \beta) \; J(\lambda, \alpha, \beta) \;.
\end{equation}
In the dual-Bracewell modes, the final signal is computed as the difference between the two null outputs, in which case, one can consider the difference of the transmission maps of these outputs multiplied by the on-sky intensity distribution. Transmission maps and phase diagrams for the interferometric outputs of the different observing modes are show in Fig.~\ref{fig:obsmodes}.

\subsubsection*{Single-Bracewell Mode}
\label{subsub:sing_brace}

The single-Bracewell (SB) mode is achieved by adjusting the phase of the input beams such that on-axis light is suppressed at the pair-wise combined outputs of the beam combiner ($m = [2,5]$; see Fig.~\ref{fig:obsmodes} top). This mode produces two independent nulls similar to the classic Bracewell nulling interferometer \cite{bracewell:1978}. 

A simplified beam combiner matrix for the interferometric outputs is,
\begin{equation}
    \mathbf{M}_{\rm{SB}}
    \propto 
    \begin{bmatrix}
      -1 &  1 &  0 &  0  \\
      -1 & -1 &  j &  j  \\
      -1 & -1 & -j & -j  \\
       0 &  0 & -1 &  1 
    \end{bmatrix} \;,
\end{equation}
where the first and last row describe input beams that are $\pi$\,rad out of phase. This expression is simplified in that it is limited to on-axis light at the central wavelength of the science band.

In this study, we are interested in the null-depth, estimated as
\begin{equation}
    \label{eq:SB0_nulldetph}
    N_{\mathrm{SB}0}(\lambda) = \frac{b}{1-b} \frac{I_2(\lambda)}{I_0(\lambda) + I_1(\lambda)}
\end{equation}
and
\begin{equation}
    \label{eq:SB1_nulldetph}
    N_{\mathrm{SB}1}(\lambda) = \frac{b}{1-b} \frac{I_5(\lambda)}{I_6(\lambda) + I_7(\lambda)} \;,
\end{equation}
where $b$ is the photometric tap ratio. We use the respective photometric channels for normalization since their flux is nominally dominated by the on-axis star and are a reasonable proxy for the typically used constructive fringe which we do not have access to with NOTT.

Key advantages of the SB mode are high throughput of off-axis light, and two simultaneous independent single-baseline measurements. The independence of the single-baselines maximally preserves spatial information. With sufficient Fourier plane coverage, it allows for strait-forward image reconstruction (with the limitation of enforced symmetry), using a process similar to regular aperture synthesis. The main disadvantage of the SB mode is the lack of any inherent error suppression, which is expected to result in the worst null-depth precision of the candidate observing modes.

\subsubsection*{Asymmetric Dual-Bracewell Mode}
\label{subsub:asym}

The asymmetric dual-Bracewell (asym) mode is achieved by adjusting the phase of the input beams such that the single-Bracewell nulls are sent to the second stage directional coupler (see Fig.~\ref{fig:combiner}) producing two complementary nulls at outputs ($m = [3,4]$). These nulls are a mixture of signal from all four input beams and contain simultaneous information from all 6 possible baselines (see Fig.~\ref{fig:obsmodes} middle). 

Transmission maps for the null outputs are generally asymmetric and produced from the combination of two single-Bracewell nulls, thus the name asymmetric dual-Bracewell mode.

A simplified beam combiner matrix for the interferometric outputs is,
\begin{equation}
    \mathbf{M}_{\rm{asym}}
    \propto 
    \begin{bmatrix}
       1 &  1 &  0 &  0  \\
       1 & -1 & -j &  j  \\
       1 & -1 &  j & -j  \\
       0 &  0 &  1 &  1 
    \end{bmatrix} \; .
\end{equation}
Note that the null outputs are complex conjugates of each other. In this way, both outputs contain correlated instrumental errors that partially cancel when the outputs are subtracted. Additionally, since each null output contains equal contributions from the individual input beams, subtracting these outputs removes independent background variations introduced by the input beams.
Differential null depth is calculated using
\begin{equation}
    \label{eq:asym_nulldepth}
    N_{\mathrm{\mathrm{asym}}}(\lambda) = \frac{b}{1-b} \left[ \frac{I_3(\lambda)}{I_0(\lambda) + I_1(\lambda)} - \frac{I_4(\lambda)}{I_6(\lambda) + I_7(\lambda)} \right] \;.
\end{equation}
The term in brackets computes the difference between the null depths associated with individual null outputs.

The key advantage of the asym mode is its expected inherent suppression of instrumental errors and background fluctuations providing the deepest and broadest nulls. However, this mode rejects symmetric signal and results in low throughput for extended targets. This can be observed by inspecting the middle transmission map in Fig.~\ref{fig:obsmodes}. This map contains complementary positive and negative regions that when multiplied by a symmetric or generally extended source will result in significant signal suppression. Additionally, the signal at the null outputs is a complicated relation of six independent baselines making it difficult to interpret the spatial information.

\subsubsection*{Symmetric Dual-Bracewell Mode}
\label{subsub:sym}

The symmetric dual-Bracewell (sym) mode is similar to the asym mode but with a relative phase shift between the two single-Bracewell nulls before the second-stage directional coupler (see Fig.~\ref{fig:obsmodes} bottom). In this case, the transmission maps for the null outputs are generally symmetric and produced from the combination of two single-Bracewell nulls, thus the name symmetric dual-Bracewell mode. 

A simplified beam combiner matrix for the interferometric outputs is
\begin{equation}
    \mathbf{M}_{\rm{sym}}
    \propto
    \begin{bmatrix}
       1 &  1 &  0 &  0  \\
       1 & -1 & -1 &  1  \\
       1 & -1 &  1 & -1  \\
       0 &  0 & -j & -j 
    \end{bmatrix} \;.
\end{equation}
In this case, the null outputs consist of purely real contributions from the input beams which results in the suppression of some instrumental error. Background fluctuations are also suppressed since the differential null-depth is calculated in the same way as for the asym mode (Eq.~\ref{eq:asym_nulldepth}), thus $N_{\mathrm{\mathrm{sym}}}(\lambda) = N_{\mathrm{\mathrm{asym}}}(\lambda)$.

The key advantage of the sym mode over the SB mode is its modest suppression of instrumental error and background fluctuations. It can be observed from the bottom transmission map of Fig.~\ref{fig:obsmodes} that there is a bias towards negative regions. As such, the sym mode has some sensitivity to symmetric and extended targets though the positive regions still reduce overall throughput.

\section{SIMULATED PERFORMANCE}
\label{sec:performance}

The Python package SCIFYsim has been developed to perform rigorous simulations of nulling interferometers \cite{laugier:2023}. After having adapted and optimized this software for the simulation of NOTT's performance in the candidate observing modes, we simulate NOTT observations for a variety of isolated stars and quantify the null-depth statistics. We focus on isolated stars as their flux dominates the expected science target's flux (i.e., planetary mass companions or HEZD), even in the null outputs. 

For the simulations in this work, the NOTT instrument has been configured with a spectral resolving power of $R=60$, a photometric tap ratio of $p=0.4$, and a detector integration time of \qty{0.1}{\second}. Furthermore, the directional couplers are modeled with chromatic amplitude and phase properties following design specifications\cite{sharma:2020}. We consider the VLTI ATs in the small array configurations; the ATs provide sufficient collecting area to meet sensitivity requirements for HEZD observations, while the choice of input array configuration has no impact on the derived noise properties.

\subsection{Null-Depth Precision}
\label{subsec:null_prec}

We assess the instrumental null-depth precision by comparing how the null depth changes with repeated measurements of the same target star in the presence of realistic noise. Variations in measured null depths result from three categories of errors:

\noindent\textbf{Instrumental Errors}: These include wavefront, tip-tilt, and piston errors. The wavefront errors are simulated by moving a static phase map over the AT apertures. The map is generated from a spectrum that models realistic residual atmospheric density fluctuations and tip-tilt jitter after wavefront correction. Piston error results from intraband dispersion due to dry-air fluctuations and interband dispersion between the fringe tracker ($K$)\cite{taras:2024} and science ($L$) bands resulting from water vapor density fluctuations. The intraband component of the piston is modeled after measured VLTI/GRAVITY instrumental piston errors\cite{lacour:2019}. The interband component is derived by modeling realistic water vapor fluctuations with an approach similar to previous work\cite{absil:2022}, and propagating these effects into a piston error. 

\noindent\textbf{Background Variations}: These are flux variations that result from the detectors seeing time varying thermal loads. The warm and cold optics of NOTT are assumed to be thermally stable with the temperature variations of the atmosphere and AT mirrors dominating this effect. We use temperature time series measurements at Paranal Observatory to generate realistic temperature fluctuations and inject these into SCIFYsim. We assume the temperatures of the atmosphere and collecting mirrors are correlated with a sensitivity coefficient of 0.8 for the mirrors (i.e., 0.8\,K change in mirror temperature for every 1\,K change in atmospheric temperature).

\noindent\textbf{Photon Noise}: This is the fundamental noise resulting from the particle-like nature of photons that scales as the square root of the number of photons collected by each detector pixel during the integration time.

\noindent Additional sources of noise include the dark current and readout noise of the detector system which are negligible compared to the other sources of error.

We find that the null-depth precision primarily depends on the stellar magnitude and is mostly independent from the stellar spectral type.
This is as expected as the magnitude of a point-like target star should be the main driver of precision in the expected contrast-limited and sensitivity-limited regimes.
For these experiments, we simulate an A5V-type star as a uniform disk and place it at various distances from the observer so as to produce samples equally spaced in $L$-band magnitudes. 

For each $L$-band magnitude, we simulate four batches of \num{3,000} individual \qty{0.1}{\second} exposures. Each batch has a unique random number generation seed, resulting in similar, but independent realizations of instrumental and background variations. This structure is modeled after our expected CAL-SCI-CAL-SCI observing strategy where we alternate between observing calibrator (CAL) and science (SCI) targets at a \qty{5}{\min} cadence (ignoring overheads). 

For each exposure, we compute the null-depth spectrum and reduce it to a single value by taking its median. Of these values, we compute the average $\mu_i$ and standard error $\sigma_i$ for each batch $i$. Then, we compute the standard error between batches $\sigma_{\mu_i}$ and the average standard error between batches $\bar{\sigma_i}$. The final precision estimate at a given $L$-band sample is
\begin{equation}
    \label{eq:err_estimate}
    \sigma_{L} = \frac{\sigma_{\mu_i} + \bar{\sigma_i}}{2} \;.
\end{equation}
For bright stars (contrast-limited regime), the variation between batches dominates and is limited by instrumental errors. For faint stars (sensitivity-limited regime), the variation within a batch dominates and is limited by photon noise and background variations.

\begin{figure}
    \centering
    \includesvg[width=0.5\linewidth]{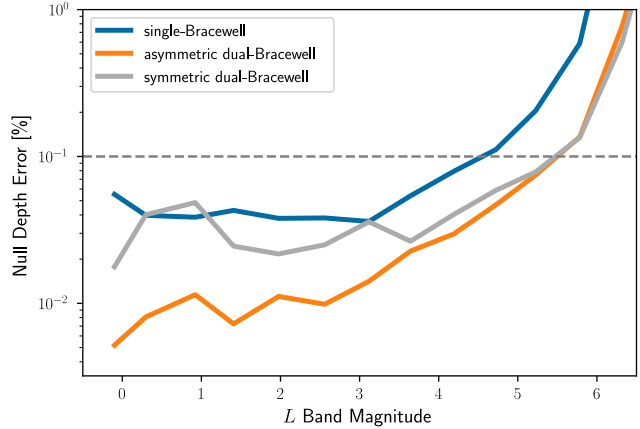}
    \caption{Null-depth precision of the different observing modes. NOTT is limited by instrumental errors for low magnitude stars and photon noise from background emission for higher magnitude stars. The dashed line indicates a reference precision used to identify the limiting magnitudes for the different modes.}
    \label{fig:null_precision}
\end{figure}

The simulated results for the null-depth precision using the candidate observing modes are shown in Fig.~\ref{fig:null_precision}. All modes show the best performance observing bright stars, which deteriorates for fainter stars. The asym mode has the best rejection of instrumental errors, visible for magnitudes $\lesssim \num{3}$. There, both SB and sym show worse performance than asym, but mutually similar. Null-depth precision degrades towards fainter stars for all modes. The SB mode degrades most rapidly due to poor background rejection while the decrease of the asym and sym modes performance is driven primarily by photon noise.

We choose a desired null-depth precision of \qty{0.1}{\percent} to define the limiting magnitude of observations. In SB mode, approximately half of the light from an extended circumstellar disk is transmitted through the normalized transmission pattern. As such, an extended disk with a \qty{1}{\percent} disk-to-star flux ratio can be measured with a SNR ratio of 5. The limiting magnitude for SB is $\sim$4.5, while the limiting magnitude for asym and sym is $\sim$5.5, although the expected throughput from extended emission is lower than for the SB mode (see next section).

\subsection{Target Null-Depth Curves}
\label{subsec:null_curves}

In the previous section we estimated the null-depth sensitivity of the candidate observing modes, which represents the uncertainty of a typical observing sequence. To better understand the amplitude of the null-depth signal, we compute noise free null-depth curves for reasonable observational scenarios using SCIFYsim.

\begin{figure}[htbp]
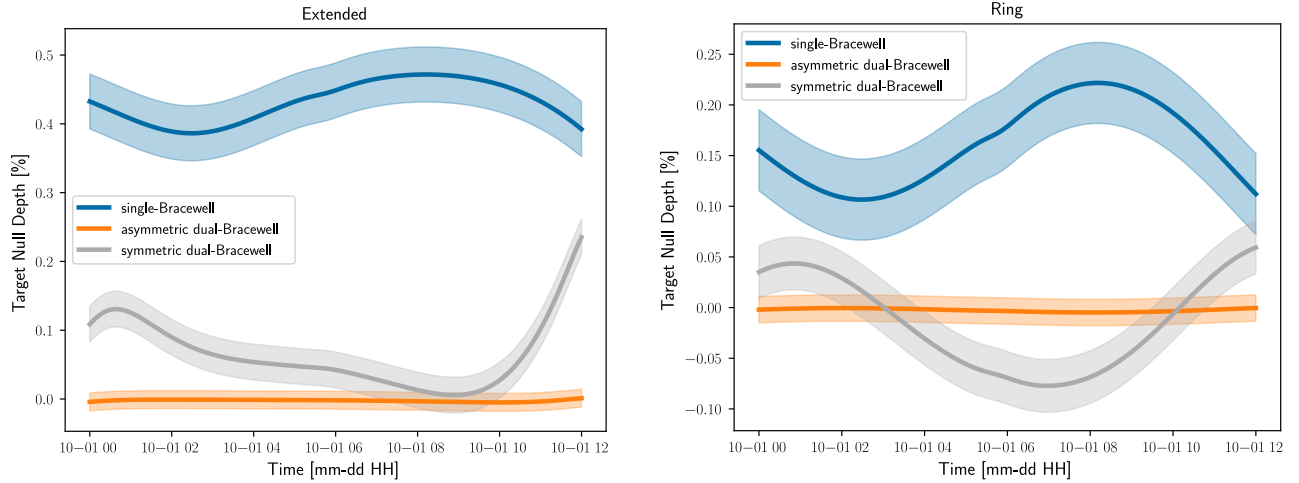

    \centering

    \begin{subfigure}[b]{0.48\textwidth}
        \centering
        \includesvg[width=\textwidth]{figures/extended_inc60.svg} 
    \end{subfigure}
    \hfill
    \begin{subfigure}[b]{0.50\textwidth}
        \centering
        \includesvg[width=\textwidth]{figures/ring_inc60.svg} 
    \end{subfigure}

    \caption{Simulated null-depth curves for extended (left) and narrow ring (right) HEZD disks around a star resembling $\tau$ Ceti. Bands enveloping the curves show the estimated error for a star with an $L$-band magnitude of \num{3}. Both disks have an inclination of 60$^\circ$ with a disk-to-star flux ratio of \qty{1}{\percent}.}
    \label{fig:disk_curves}
\end{figure}

The target star is modeled after the star $\tau$ Ceti (radius, effective temperature, and sky position) being observed from Paranal Observatory for \qty{12}{\hour} starting at midnight UTC (8 pm local) October 1, 2026. Tau Ceti has an $L$-band magnitude of $\sim$3 which determines the null-depth precision of NOTT in the different observing modes (see Fig.~\ref{fig:null_precision}).
We consider two dust geometries, both having a 60$^\circ$ inclination with a dust-to-star flux ratio of \qty{1}{\percent}. The dust model is based on previous work\cite{kennedy:2015}, where the dust is assumed to be optically thin and emit as a blackbody. The surface density and temperature of the dust decreases radially away from the host star according to a power law distribution. For one model, the dust distribution is extended with an inner radius at the dust sublimation temperature (\qty{1500}{K}) and extends into the habitable zone (\qty{300}{K}). For the other, it is distributed in a narrow ring extending from the sublimation radius to a distance \qty{10}{\percent} larger. The simulated null-depth curves for these models are shown in Fig.~\ref{fig:disk_curves} 

Although the asym mode has the smallest uncertainty, it produces negligible null-depth signal. The SB mode has the most advantageous SNR for these disk models while delivering two measurements (one for each baseline; only one shown in the figure), providing more spatial information and higher confidence detections. However, for fainter stars, the sym mode would deliver superior SNR (for this specific dust geometry) due to the more slowly degrading null-depth precision (see Fig.~\ref{fig:null_precision}).

\begin{figure}
    \centering
    \includesvg[width=0.5\linewidth]{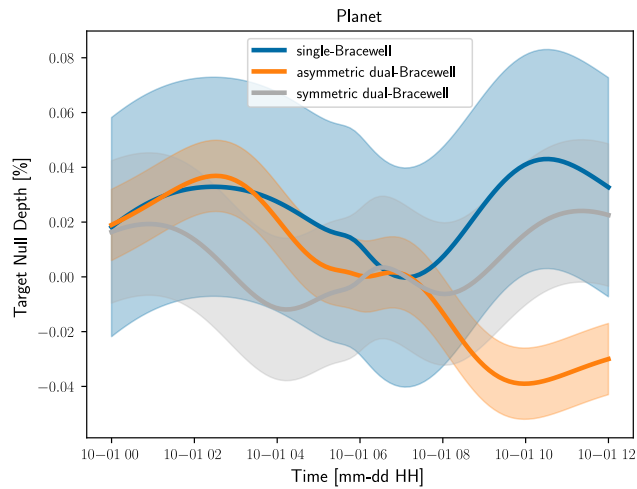}
    \caption{Simulated null-depth curves for a synthetic hot Jupiter around a star resembling $\tau$~Ceti with an $L$-band magnitude of \num{3}. Bands enveloping the curves show the estimated error for a star with an $L$-band magnitude of \num{3}.}
    \label{fig:planet_curves}
\end{figure}

Additionally, we simulated the system without HEZD, but with an exoplanetary companion with a radius of 1.5 times that of Jupiter and a temperature of \qty{900}{K} (i.e., a hot Jupiter), being located \qty{80}{mas} from the host star. The simulated null-depth curves are shown in Fig.~\ref{fig:planet_curves}. In this case, the amplitudes of the null-depth curves for the different observing modes are similar but the SNR of asym mode is significantly higher, illustrating its utility for the detection of point-like features such as stellar or substellar companions or clumpy disk features.

This study shows that the candidate observing modes have different trade-offs in terms of amplitude and achievable null-depth precision. Which mode should be used depends on the specific observational scenario. The trade-offs and recommended targets are summarized in Table.~\ref{tab:bracewell_comparison}.

\begin{table}[htbp]
    \centering
    \caption{Comparison of NOTT observing modes.}
    \label{tab:bracewell_comparison}
    \small 
    \begin{tabularx}{\textwidth}{>{\raggedright\arraybackslash}p{2.8cm} L L L}
        \toprule
        \textbf{Mode} & \textbf{Advantages} & \textbf{Disadvantages} & \textbf{Recommended Targets} \\
        \midrule
        
        \textbf{Single-Bracewell} & 
        \textbullet~Preserves spatial information \newline \textbullet~High throughput & 
        \textbullet~Poor error suppression & 
        \textbullet~Disks and extended targets around $\text{mag} \le 4$ stars \\
        \addlinespace
        
        \textbf{Asymmetric dual-Bracewell} & 
        \textbullet~Best error suppression & 
        \textbullet~Insensitive to symmetric targets \newline \textbullet~Low throughput & 
        \textbullet~Exoplanets and compact structures \\
        \addlinespace
        
        \textbf{Symmetric dual-Bracewell} & 
        \textbullet~Sensitive to symmetric targets \newline \textbullet~Background suppression & 
        \textbullet~Modest instrumental error suppression \newline \textbullet~Low throughput & 
        \textbullet~Disks and extended structures around $4 < \text{mag} < 5$ stars \\
        
        \bottomrule
    \end{tabularx}
\end{table}

\section{CONCLUSION}
\label{sec:conclusion}

We have studied three distinct observing modes for the NOTT nulling interferometer, each presenting trade-offs among throughput, sensitivity to various target geometries (specifically symmetry), and error suppression. The single-Bracewell mode, using a pairwise telescope combination, produces the poorest error suppression and is constrained by the brightest limiting magnitude. In comparison, the symmetric dual-Bracewell mode, a four-telescope combination engineered for symmetric emission, achieves stronger error suppression alongside fainter limiting magnitudes. Although both modes share similar overall sensitivity to symmetric HEZD emission, their observations probe complementary spatial scales. On the other hand, while the asymmetric dual-Bracewell mode provides the highest degree of error suppression, it inherently suppresses symmetric astrophysical signals, rendering it incapable of detecting the simulated HEZD emission. However, this trade-off reverses when targeting asymmetric point sources like hot Jupiters. Ultimately, combining data from all three modes provides a powerful approach for detecting and characterizing HEZD, constraining azimuthal asymmetries, and identifying potential false positives from binary companion stars.

\appendix    

\acknowledgments 
 
This work is supported by the National Aeronautics and Space Administration
(NASA) through grant \\ 80NSSC23K1473 (J.P.S., S.E., T.A.S.).
SCIFY has received funding from the European Research Council (ERC); Award no. CoG – 866070 under the European Union’s Horizon 2020 research and innovation program.

\bibliography{bibliography} 

\begin{thebibliography}{10}

\bibitem{kral:2017}
{Kral}, Q., {Krivov}, A.~V., {Defr{\`e}re}, D., {van Lieshout}, R., {Bonsor}, A., {Augereau}, J.-C., {Th{\'e}bault}, P., {Ertel}, S., {Lebreton}, J., and {Absil}, O., ``{Exozodiacal clouds: hot and warm dust around main sequence stars},'' {\em The Astronomical Review}~{\bf 13},  69--111 (Apr. 2017).

\bibitem{ertel:2025}
{Ertel}, S., {Pearce}, T.~D., {Debes}, J.~H., {Faramaz}, V.~C., {Danchi}, W.~C., {Anche}, R.~M., {Defr{\`e}re}, D., {Hasegawa}, Y., {Hom}, J., {Kirchschlager}, F., {Rebollido}, I., {Rousseau}, H., {Scott}, J., {Stapelfeldt}, K., and {Stuber}, T.~A., ``{Review and Prospects of Hot Exozodiacal Dust Research For Future Exo-Earth Direct Imaging Missions},'' {\em \pasp}~{\bf 137},  031001 (Mar. 2025).

\bibitem{absil:2013}
{Absil}, O., {Defr{\`e}re}, D., {Coud{\'e} du Foresto}, V., {di Folco}, E., {M{\'e}rand}, A., {Augereau}, J.~C., {Ertel}, S., {Hanot}, C., {Kervella}, P., {Mollier}, B., {Scott}, N., {Che}, X., {Monnier}, J.~D., {Thureau}, N., {Tuthill}, P.~G., {ten Brummelaar}, T.~A., {McAlister}, H.~A., {Sturmann}, J., {Sturmann}, L., and {Turner}, N., ``{A near-infrared interferometric survey of debris-disc stars. III. First statistics based on 42 stars observed with CHARA/FLUOR},'' {\em \aap}~{\bf 555},  A104 (July 2013).

\bibitem{nunez:2017}
{Nu{\~n}ez}, P.~D., {Scott}, N.~J., {Mennesson}, B., {Absil}, O., {Augereau}, J.~C., {Bryden}, G., {ten Brummelaar}, T., {Ertel}, S., {Coud{\'e} du Foresto}, V., {Ridgway}, S.~T., {Sturmann}, J., {Sturmann}, L., {Turner}, N.~J., and {Turner}, N.~H., ``{A near-infrared interferometric survey of debris-disc stars. VI. Extending the exozodiacal light survey with CHARA/JouFLU},'' {\em \aap}~{\bf 608},  A113 (Dec. 2017).

\bibitem{ertel:2014}
{Ertel}, S., {Absil}, O., {Defr{\`e}re}, D., {Le Bouquin}, J.~B., {Augereau}, J.~C., {Marion}, L., {Blind}, N., {Bonsor}, A., {Bryden}, G., {Lebreton}, J., and {Milli}, J., ``{A near-infrared interferometric survey of debris-disk stars. IV. An unbiased sample of 92 southern stars observed in H band with VLTI/PIONIER},'' {\em \aap}~{\bf 570},  A128 (Oct. 2014).

\bibitem{absil:2021}
{Absil}, O., {Marion}, L., {Ertel}, S., {Defr{\`e}re}, D., {Kennedy}, G.~M., {Romagnolo}, A., {Le Bouquin}, J.~B., {Christiaens}, V., {Milli}, J., {Bonsor}, A., {Olofsson}, J., {Su}, K.~Y.~L., and {Augereau}, J.~C., ``{A near-infrared interferometric survey of debris-disk stars. VII. The hot-to-warm dust connection},'' {\em \aap}~{\bf 651},  A45 (July 2021).

\bibitem{absil:2006}
{Absil}, O., {di Folco}, E., {M{\'e}rand}, A., {Augereau}, J.~C., {Coud{\'e} du Foresto}, V., {Aufdenberg}, J.~P., {Kervella}, P., {Ridgway}, S.~T., {Berger}, D.~H., {ten Brummelaar}, T.~A., {Sturmann}, J., {Sturmann}, L., {Turner}, N.~H., and {McAlister}, H.~A., ``{Circumstellar material in the Vega inner system revealed by CHARA/FLUOR},'' {\em \aap}~{\bf 452},  237--244 (June 2006).

\bibitem{akeson:2009}
{Akeson}, R.~L., {Ciardi}, D.~R., {Millan-Gabet}, R., {Merand}, A., {di Folco}, E., {Monnier}, J.~D., {Beichman}, C.~A., {Absil}, O., {Aufdenberg}, J., {McAlister}, H., {ten Brummelaar}, T., {Sturmann}, J., {Sturmann}, L., and {Turner}, N., ``{Dust in the inner regions of debris disks around a stars},'' {\em \apj}~{\bf 691},  1896--1908 (Feb. 2009).

\bibitem{kirchschlager:2017}
{Kirchschlager}, F., {Wolf}, S., {Krivov}, A.~V., {Mutschke}, H., and {Brunngr{\"a}ber}, R., ``{Constraints on the structure of hot exozodiacal dust belts},'' {\em \mnras}~{\bf 467},  1614--1626 (May 2017).

\bibitem{stuber:2023b}
{Stuber}, T.~A., {Kirchschlager}, F., {Pearce}, T.~D., {Ertel}, S., {Krivov}, A.~V., and {Wolf}, S., ``{How much large dust could be present in hot exozodiacal dust systems?},'' {\em \aap}~{\bf 678},  A121 (Oct. 2023).

\bibitem{pearce:2020}
{Pearce}, T.~D., {Krivov}, A.~V., and {Booth}, M., ``{Gas trapping of hot dust around main-sequence stars},'' {\em \mnras}~{\bf 498},  2798--2813 (Oct. 2020).

\bibitem{pearce:2022}
{Pearce}, T.~D., {Kirchschlager}, F., {Rouill{\'e}}, G., {Ertel}, S., {Bensberg}, A., {Krivov}, A.~V., {Booth}, M., {Wolf}, S., and {Augereau}, J.-C., ``{Hot exozodis: cometary supply without trapping is unlikely to be the mechanism},'' {\em \mnras}~{\bf 517},  1436--1451 (Nov. 2022).

\bibitem{quanz:2022}
{Quanz}, S.~P., {Ottiger}, M., {Fontanet}, E., {Kammerer}, J., {Menti}, F., {Dannert}, F., {Gheorghe}, A., {Absil}, O., {Airapetian}, V.~S., {Alei}, E., {Allart}, R., {Angerhausen}, D., {Blumenthal}, S., {Buchhave}, L.~A., {Cabrera}, J., {Carri{\'o}n-Gonz{\'a}lez}, {\'O}., {Chauvin}, G., {Danchi}, W.~C., {Dandumont}, C., {Defr{\'e}re}, D., {Dorn}, C., {Ehrenreich}, D., {Ertel}, S., {Fridlund}, M., {Garc{\'\i}a Mu{\~n}oz}, A., {Gasc{\'o}n}, C., {Girard}, J.~H., {Glauser}, A., {Grenfell}, J.~L., {Guidi}, G., {Hagelberg}, J., {Helled}, R., {Ireland}, M.~J., {Janson}, M., {Kopparapu}, R.~K., {Korth}, J., {Kozakis}, T., {Kraus}, S., {L{\'e}ger}, A., {Leedj{\"a}rv}, L., {Lichtenberg}, T., {Lillo-Box}, J., {Linz}, H., {Liseau}, R., {Loicq}, J., {Mahendra}, V., {Malbet}, F., {Mathew}, J., {Mennesson}, B., {Meyer}, M.~R., {Mishra}, L., {Molaverdikhani}, K., {Noack}, L., {Oza}, A.~V., {Pall{\'e}}, E., {Parviainen}, H., {Quirrenbach}, A., {Rauer}, H., {Ribas}, I., {Rice}, M., {Romagnolo}, A., {Rugheimer}, S.,
  {Schwieterman}, E.~W., {Serabyn}, E., {Sharma}, S., {Stassun}, K.~G., {Szul{\'a}gyi}, J., {Wang}, H.~S., {Wunderlich}, F., {Wyatt}, M.~C., and {LIFE Collaboration}, ``{Large Interferometer For Exoplanets (LIFE). I. Improved exoplanet detection yield estimates for a large mid-infrared space-interferometer mission},'' {\em \aap}~{\bf 664},  A21 (Aug. 2022).

\bibitem{defrere:2018a}
{Defr{\`e}re}, D., {Ireland}, M., {Absil}, O., {Berger}, J.~P., {Danchi}, W.~C., {Ertel}, S., {Gallenne}, A., {H{\'e}nault}, F., {Hinz}, P., {Huby}, E., {Kraus}, S., {Labadie}, L., {Le Bouquin}, J.~B., {Martin}, G., {Matter}, A., {Mennesson}, B., {M{\'e}rand}, A., {Minardi}, S., {Monnier}, J.~D., {Norris}, B., {Orban de Xivry}, G., {Pedretti}, E., {Pott}, J.~U., {Reggiani}, M., {Serabyn}, E., {Surdej}, J., {Tristram}, K.~R.~W., and {Woillez}, J., ``\emph{Hi-5: a potential high-contrast thermal near-infrared imager for the VLTI},'' in [{\em Optical and Infrared Interferometry and Imaging VI}{\nolinebreak\hspace{0.1em}]},  {Creech-Eakman}, M.~J., {Tuthill}, P.~G., and {M{\'e}rand}, A., eds., {\em \procspie} {\bf 10701},  107010U (July 2018).

\bibitem{defrere:2018b}
{Defr{\`e}re}, D., {Absil}, O., {Berger}, J.~P., {Boulet}, T., {Danchi}, W.~C., {Ertel}, S., {Gallenne}, A., {H{\'e}nault}, F., {Hinz}, P., {Huby}, E., {Ireland}, M., {Kraus}, S., {Labadie}, L., {Le Bouquin}, J.~B., {Martin}, G., {Matter}, A., {M{\'e}rand}, A., {Mennesson}, B., {Minardi}, S., {Monnier}, J.~D., {Norris}, B., {Orban de Xivry}, G., {Pedretti}, E., {Pott}, J.~U., {Reggiani}, M., {Serabyn}, E., {Surdej}, J., {Tristram}, K.~R.~W., and {Woillez}, J., ``{The path towards high-contrast imaging with the VLTI: the Hi-5 project},'' {\em Experimental Astronomy}~{\bf 46},  475--495 (Dec. 2018).

\bibitem{defrere:2022}
{Defr{\`e}re}, D., {Bigioli}, A., {Dandumont}, C., {Garreau}, G., {Laugier}, R., {Martinod}, M.-A., {Absil}, O., {Berger}, J.-P., {Bouzerand}, E., {Courtney-Barrer}, B., {Emsenhuber}, A., {Ertel}, S., {Gagne}, J., {Glauser}, A., {Gross}, S., {Ireland}, M.~J., {Kenchington}, H.-D., {Kluska}, J., {Kraus}, S., {Labadie}, L., {Laborde}, V., {L{\'e}ger}, A., {Leisenring}, J., {Loicq}, J., {Martin}, G., {Morren}, J., {Matter}, A., {Mazzoli}, A., {Missiaen}, K., {Muhammad}, S., {Ollivier}, M., {Raskin}, G., {Rousseau}, H., {Sanny}, A., {Verlinden}, S., {Vandenbussche}, B., and {Woillez}, J., ``\emph{L-band nulling interferometry at the VLTI with Asgard/Hi-5: status and plans},'' in [{\em Optical and Infrared Interferometry and Imaging VIII}{\nolinebreak\hspace{0.1em}]},  {M{\'e}rand}, A., {Sallum}, S., and {Sanchez-Bermudez}, J., eds., {\em \procspie} {\bf 12183},  121830H (Aug. 2022).

\bibitem{defrere:2024}
{Defr{\`e}re}, D., {Laugier}, R., {Martinod}, M.-A., {Garreau}, G., {Missiaen}, K., {Salman}, M., {Raskin}, G., {Dandumont}, C., {Ertel}, S., {Ireland}, M.~J., {Kraus}, S., {Labadie}, L., {Mazzolli}, A., {Medgyesi}, G., {Sanny}, A., {Absil}, O., {{\'A}br{\'a}ham}, P., {Berger}, J.-P., {Bonduelle}, M., {Bigioli}, A., {Bouzerand}, E., {Carter}, J., {Cvetojevic}, N., {Courtney-Barrer}, B., {Glauser}, A.~M., {Gross}, S., {Haubois}, X., {James}, N., {Joo}, A.~P., {Lagarde}, S., {L{\'e}ger}, A., {Leisenring}, J., {Loicq}, J., {Martin}, G., {Martinache}, F., {Mezo}, G., {Morel}, S., {Morren}, J., {Ollivier}, M., {Robertson}, G., {Rousseau}, H., {Schofield}, W., {Schuhler}, N., {Taras}, A., {Vandenbussche}, B., and {Woillez}, J., ``{L-band nulling interferometry at the VLTI with Asgard/NOTT: status and plans},'' in [{\em Optical and Infrared Interferometry and Imaging IX}{\nolinebreak\hspace{0.1em}]},  {Kammerer}, J., {Sallum}, S., and {Sanchez-Bermudez}, J., eds., {\em \procspie} {\bf 13095},  130950F (Aug. 2024).

\bibitem{laugier:2023}
{Laugier}, R., {Defr{\`e}re}, D., {Woillez}, J., {Courtney-Barrer}, B., {Dannert}, F.~A., {Matter}, A., {Dandumont}, C., {Gross}, S., {Absil}, O., {Bigioli}, A., {Garreau}, G., {Labadie}, L., {Loicq}, J., {Martinod}, M.-A., {Mazzoli}, A., {Raskin}, G., and {Sanny}, A., ``{Asgard/NOTT: L-band nulling interferometry at the VLTI. I. Simulating the expected high-contrast performance},'' {\em \aap}~{\bf 671},  A110 (Mar. 2023).

\bibitem{garreau:2022}
{Garreau}, G., {Bigioli}, A., {Raskin}, G., {Berger}, J.-P., {Dandumont}, C., {Kenchington Goldsmith}, H.-D., {Gross}, S., {Ireland}, M., {Labadie}, L., {Laugier}, R., {Loicq}, J., {Madden}, S., {Martin}, G., {Mazzoli}, A., {Morren}, J., {Shao}, H., {Yan}, K., and {Defr{\`e}re}, D., ``{Design of the VLTI/Hi-5 light injection system},'' in [{\em Optical and Infrared Interferometry and Imaging VIII}{\nolinebreak\hspace{0.1em}]},  {M{\'e}rand}, A., {Sallum}, S., and {Sanchez-Bermudez}, J., eds., {\em \procspie} {\bf 12183},  1218320 (Aug. 2022).

\bibitem{garreau:2024a}
{Garreau}, G., {Bigioli}, A., {Laugier}, R., {Raskin}, G., {Morren}, J., {Berger}, J.-P., {Dandumont}, C., {Goldsmith}, H.-D.~K., {Gross}, S., {Ireland}, M., {Labadie}, L., {Loicq}, J., {Madden}, S., {Martin}, G., {Martinod}, M.-A., {Mazzoli}, A., {Sanny}, A., {Shao}, H., {Yan}, K., and {Defr{\`e}re}, D., ``{Asgard/NOTT: L-band nulling interferometry at the VLTI. II. Warm optical design and injection system},'' {\em Journal of Astronomical Telescopes, Instruments, and Systems}~{\bf 10},  015002 (Jan. 2024).

\bibitem{garreau:2024b}
{Garreau}, G., {Bigioli}, A., {Laugier}, R., {La Torre}, B., {Martinod}, M.-A., {Missiaen}, K., {Morren}, J., {Raskin}, G., {Salman}, M., {Gross}, S., {Ireland}, M., {Jo{\'o}}, A.~P., {Labadie}, L., {Madden}, S., {Mazzoli}, A., {Medgyesi}, G., {Sanny}, A., {Taras}, A., {Vandenbussche}, B., and {Defr{\`e}re}, D., ``{Asgard/NOTT: first lab assembly and experimental results},'' in [{\em Optical and Infrared Interferometry and Imaging IX}{\nolinebreak\hspace{0.1em}]},  {Kammerer}, J., {Sallum}, S., and {Sanchez-Bermudez}, J., eds., {\em \procspie} {\bf 13095},  130950P (Aug. 2024).

\bibitem{sanny:2026}
{Sanny}, A., {Labadie}, L., {Gross}, S., {Barjot}, K., {Laugier}, R., {Garreau}, G., {Martinod}, M.-A., {Defr{\`e}re}, D., and {Withford}, M.~J., ``{Asgard/NOTT: L-band nulling interferometry at the VLTI: III. The mid-infrared integrated optics beam combiner for NOTT},'' {\em \aap}~{\bf 705},  A37 (Jan. 2026).

\bibitem{martinod:2023}
{Martinod}, M.-A., {Defr{\`e}re}, D., {Ireland}, M., {Kraus}, S., {Martinache}, F., {Tuthill}, P., {Bigioli}, A., {Bouzerand}, E., {Bryant}, J., {Chhabra}, S., {Courtney-Barrer}, B., {Crous}, F., {Cvetojevic}, N., {Dandumont}, C., {Ertel}, S., {Gardner}, T., {Garreau}, G., {Glauser}, A.~M., {Labadie}, L., {Lagadec}, T., {Laugier}, R., {Mazzoli}, A., {Mortimer}, D., {Norris}, B., {Raskin}, G., {Robertson}, G., {Sanny}, A., and {Taras}, A., ``{High-angular resolution and high contrast observations from Y to L band at the Very Large Telescope Interferometer with the Asgard Instrumental suite},'' {\em Journal of Astronomical Telescopes, Instruments, and Systems}~{\bf 9},  025007 (Apr. 2023).

\bibitem{bracewell:1978}
{Bracewell}, R.~N., ``{Detecting nonsolar planets by spinning infrared interferometer},'' {\em \nat}~{\bf 274},  780--781 (Aug. 1978).

\bibitem{haubois:2020}
{Haubois}, X., {Abuter}, R., {Aller-Carpentier}, E., {Alonso}, J., {Beltran}, J., {Berger}, J.-P., {Bourget}, P., {Bristow}, P., {Caniguante}, L., {Chazelas}, B., {Cid}, C., {Conzelmann}, R., {Cortes}, A., {Darr{\'e}}, P., {Delboulb{\'e}}, A., {Delplancke-Str{\"o}bele}, F., {Del Valle}, D., {Dembet}, R., {Donoso}, R., {Dupuy}, C., {Egner}, S., {Eisenhauer}, F., {Faundez}, L., {Fuenteseca}, E., {Frahm}, R., {Gaytan}, D., {Gil}, J.~P., {Glindemann}, A., {Gont{\'e}}, F., {Gonzales}, J., {Guajardo}, P., {Guerlet}, T., {Guieu}, S., {Gutierrez}, P., {Haguenauer}, P., {van der Heyden}, P., {Huber}, S., {Hubin}, N., {Hummel}, C., {Jochum}, L., {Jocou}, L., {Kirchbauer}, J.-P., {Kolb}, J., {Kosmalski}, J., {Krempl}, P., {Lacour}, S., {Le Bouquin}, J.-B., {Leclercq}, J., {Lizon}, J.~L., {Lopez}, B., {Magnard}, Y., {Meilland}, A., {Meister}, A., {M{\'e}rand}, A., {Mieske}, S., {Moulin}, T., {Osorio}, J., {Ott}, J., {Paladini}, C., {Pallanca}, L., {Pavez}, M., {Pasquini}, L., {Pelluet}, C., {Percheron}, I., {Pettazzi},
  L., {Pino}, A., {Poupar}, S., {Ram{\'\i}rez}, A., {Reyes}, J., {Riquelme}, M., {Rivinius}, T., {Rochat}, S., {Salgado}, F., {Sch{\"o}ller}, M., {Schuhler}, N., {Shchekaturov}, P., {Stephan}, C., {Suarez}, M., {Smette}, A., {Tamblay}, R., {Tapia}, M., {Tristram}, K., {Valdes}, G., {Verinaud}, C., {Wittkowski}, M., {Woillez}, J., and {Zins}, G., ``{VLTI status update: tapping into a powerful second-generation instrumentation},'' in [{\em Optical and Infrared Interferometry and Imaging VII}{\nolinebreak\hspace{0.1em}]},  {Tuthill}, P.~G., {M{\'e}rand}, A., and {Sallum}, S., eds., {\em \procspie} {\bf 11446},  1144606 (Dec. 2020).

\bibitem{dandumont:2022}
{Dandumont}, C., {Laugier}, R., {Emsenhuber}, A., {Gagne}, J., {Absil}, O., {Bigioli}, A., {Bonavita}, M., {Garreau}, G., {Ireland}, M., {Martinod}, M.-A., {Loicq}, J., and {Defr{\`e}re}, D., ``{VLTI/Hi-5: detection yield predictions for young giant exoplanets},'' in [{\em Optical and Infrared Interferometry and Imaging VIII}{\nolinebreak\hspace{0.1em}]},  {M{\'e}rand}, A., {Sallum}, S., and {Sanchez-Bermudez}, J., eds., {\em \procspie} {\bf 12183},  1218327 (Aug. 2022).

\bibitem{ertel:2018a}
{Ertel}, S., {Defr{\`e}re}, D., {Hinz}, P., {Mennesson}, B., {Kennedy}, G.~M., {Danchi}, W.~C., {Gelino}, C., {Hill}, J.~M., {Hoffmann}, W.~F., {Rieke}, G., {Shannon}, A., {Spalding}, E., {Stone}, J.~M., {Vaz}, A., {Weinberger}, A.~J., {Willems}, P., {Absil}, O., {Arbo}, P., {Bailey}, V.~P., {Beichman}, C., {Bryden}, G., {Downey}, E.~C., {Durney}, O., {Esposito}, S., {Gaspar}, A., {Grenz}, P., {Haniff}, C.~A., {Leisenring}, J.~M., {Marion}, L., {McMahon}, T.~J., {Millan-Gabet}, R., {Montoya}, M., {Morzinski}, K.~M., {Pinna}, E., {Power}, J., {Puglisi}, A., {Roberge}, A., {Serabyn}, E., {Skemer}, A.~J., {Stapelfeldt}, K., {Su}, K.~Y.~L., {Vaitheeswaran}, V., and {Wyatt}, M.~C., ``{The HOSTS Survey{\textemdash}Exozodiacal Dust Measurements for 30 Stars},'' {\em \aj}~{\bf 155},  194 (May 2018).

\bibitem{ertel:2020a}
{Ertel}, S., {Defr{\`e}re}, D., {Hinz}, P., {Mennesson}, B., {Kennedy}, G.~M., {Danchi}, W.~C., {Gelino}, C., {Hill}, J.~M., {Hoffmann}, W.~F., {Mazoyer}, J., {Rieke}, G., {Shannon}, A., {Stapelfeldt}, K., {Spalding}, E., {Stone}, J.~M., {Vaz}, A., {Weinberger}, A.~J., {Willems}, P., {Absil}, O., {Arbo}, P., {Bailey}, V.~P., {Beichman}, C., {Bryden}, G., {Downey}, E.~C., {Durney}, O., {Esposito}, S., {Gaspar}, A., {Grenz}, P., {Haniff}, C.~A., {Leisenring}, J.~M., {Marion}, L., {McMahon}, T.~J., {Millan-Gabet}, R., {Montoya}, M., {Morzinski}, K.~M., {Perera}, S., {Pinna}, E., {Pott}, J.~U., {Power}, J., {Puglisi}, A., {Roberge}, A., {Serabyn}, E., {Skemer}, A.~J., {Su}, K.~Y.~L., {Vaitheeswaran}, V., and {Wyatt}, M.~C., ``{The HOSTS Survey for Exozodiacal Dust: Observational Results from the Complete Survey},'' {\em \aj}~{\bf 159},  177 (Apr. 2020).

\bibitem{kirchschlager:2020}
{Kirchschlager}, F., {Ertel}, S., {Wolf}, S., {Matter}, A., and {Krivov}, A.~V., ``{First L band detection of hot exozodiacal dust with VLTI/MATISSE},'' {\em \mnras}~{\bf 499},  L47--L52 (Dec. 2020).

\bibitem{ollmann:2025}
{Ollmann}, K., {Kirchschlager}, F., {Stuber}, T.~A., {Tsishchankava}, K., {Matter}, A., {Ertel}, S., {Pearce}, T.~D., {Krivov}, A.~V., and {Wolf}, S., ``{Hot exozodiacal dust around Fomalhaut: The MATISSE perspective},'' {\em \aap}~{\bf 699},  A144 (July 2025).

\bibitem{priolet:inpress}
{Priolet}, P., {Augereau}, J.-C., {Milli}, J., {Matter}, A., {Lopez}, B., {Beust}, H., {Varga}, J., {Boley}, P., {Danchi}, W.-C., {Foteini}, L., {Henning}, T., {Houllé}, M., {Letessier}, M., {Millour}, F., {Scigliuto}, J., {Weigelt}, G., {Wolf}, S., {Absil}, O., {Berger}, J.-P., {Bouarour}, Y.-I., {Bourdarot}, G., {Defrère}, D., {Desgrange}, C., {Le Bouquin}, J.-B., {Iglesias}, D., {Stuber}, T.~A., {Berio}, P., {Bettonvil}, F., {Cruzalèbes}, P., {Heininger}, M., {Isbell}, J.~W., {Lagarde}, S., {Meilland}, A., {Petrov}, R., {Robbe-Dubois}, S., {MATISSE Collaboration}, and {NAOMI Collaboration}, ``{VLTI/MATISSE observations of the hot dust around $\beta$ Pictoris. Observing faint objects with MATISSE},'' {\em \aap}  (in Press).

\bibitem{stuber:2026a}
{Stuber}, T.~A., {M{\'e}rand}, A., {Kirchschlager}, F., {Wolf}, S., {Weible}, G., {Absil}, O., {Pearce}, T.~D., {Garreau}, G., {Augereau}, J.-C., {Danchi}, W.~C., {Defr{\`e}re}, D., {Faramaz-Gorka}, V., {Isbell}, J.~W., {Kobus}, J., {Krivov}, A.~V., {Laugier}, R., {Ollmann}, K., {Petrov}, R.~G., {Priolet}, P., {Scott}, J.~P., {Tsishchankava}, K., and {Ertel}, S., ``{Interferometric Detection and Orbit Modeling of the Subcomponent in the Hot-dust System {\ensuremath{\kappa}} Tuc A: A Low-mass Star on an Eccentric Orbit in a Hierarchical-quintuple System},'' {\em \aj}~{\bf 171},  1 (Jan. 2026).

\bibitem{tsishchankava:2026}
{Tsishchankava}, K., {Kirchschlager}, F., {Krieger}, A., {Stuber}, T.~A., and {Wolf}, S., ``{Close-in faint companions mimicking interferometric hot exozodiacal dust observations},'' {\em \aap}~{\bf 705},  A211 (Jan. 2026).

\bibitem{garreau:2026}
{Garreau}, G., {Defr{\`e}re}, D., {Laugier}, R., {Chingaipe}, P., {Martinod}, M.-A., {Mattheussen}, T., {Missiaen}, K., {Morren}, J., {Raskin}, G., {Salman}, M., {Verstraeten}, W., {Bigioli}, A., {Ertel}, S., {Gross}, S., {Haubois}, X., {Ireland}, M., {Jo{\'o}}, A.~P., {Kraus}, S., {Labadie}, L., {Madden}, S., {Martinache}, F., {Mazzoli}, A., {Medgyesi}, G., {Sanny}, A., {Schuhler}, N., {Scott}, J.~P., {Stuber}, T.~A., and {Vandenbussche}, B., ``{Asgard/NOTT: Cryogenic characterization of the mid-infrared chip},'' {\em arXiv e-prints} ,  arXiv:2607.20191 (July 2026).

\bibitem{Hall:2011}
{Hall}, D. N.~B., ``{The Development And Use Of The HAWAII 2RG Array And SIDECAR ASIC For 1 - 5 Micron IR Observations With A Preview Of The Coming HAWAII 4RG-15.},'' in [{\em American Astronomical Society Meeting Abstracts \#217}{\nolinebreak\hspace{0.1em}]},  {\em American Astronomical Society Meeting Abstracts} {\bf 217},  425.07 (Jan. 2011).

\bibitem{sharma:2020}
{Sharma}, T.~K., {Labadie}, L., {Strixner}, D., {Gross}, S., {Gretzinger}, T., {Quanz}, S., {Defr{\`e}re}, D., {Ahmed}, S., and {Withford}, M.~J., ``{Towards the development of mid-infrared integrated optics in the renewed context of high-contrast interferometry},'' in [{\em Optical and Infrared Interferometry and Imaging VII}{\nolinebreak\hspace{0.1em}]},  {Tuthill}, P.~G., {M{\'e}rand}, A., and {Sallum}, S., eds., {\em \procspie} {\bf 11446},  1144618 (Dec. 2020).

\bibitem{taras:2024}
{Taras}, A.~K., {Robertson}, J.~G., {Allouche}, F., {Courtney-Barrer}, B., {Carter}, J., {Crous}, F., {Cvetojevic}, N., {Ireland}, M., {Lagarde}, S., {Martinache}, F., {McGinness}, G., {N'Diaye}, M., {Robbe-Dubois}, S., and {Tuthill}, P., ``{Heimdallr, Baldr, and Solarstein: designing the next generation of VLTI instruments in the Asgard suite},'' {\em \ao}~{\bf 63},  D41 (May 2024).

\bibitem{lacour:2019}
{Lacour}, S., {Dembet}, R., {Abuter}, R., {F{\'e}dou}, P., {Perrin}, G., {Choquet}, {\'E}., {Pfuhl}, O., {Eisenhauer}, F., {Woillez}, J., {Cassaing}, F., {Wieprecht}, E., {Ott}, T., {Wiezorrek}, E., {Tristram}, K.~R.~W., {Wolff}, B., {Ram{\'\i}rez}, A., {Haubois}, X., {Perraut}, K., {Straubmeier}, C., {Brandner}, W., and {Amorim}, A., ``{The GRAVITY fringe tracker},'' {\em \aap}~{\bf 624},  A99 (Apr. 2019).

\bibitem{absil:2022}
{Absil}, O., {Delacroix}, C., {Orban de Xivry}, G., {Pathak}, P., {Willson}, M., {Berio}, P., {van Boekel}, R., {Matter}, A., {Defr{\`e}re}, D., {Burtscher}, L., {Woillez}, J., and {Brandl}, B., ``{Impact of water vapor seeing on mid-infrared high-contrast imaging at ELT scale},'' in [{\em Adaptive Optics Systems VIII}{\nolinebreak\hspace{0.1em}]},  {Schreiber}, L., {Schmidt}, D., and {Vernet}, E., eds., {\em \procspie} {\bf 12185},  1218511 (Aug. 2022).

\bibitem{kennedy:2015}
{Kennedy}, G.~M., {Wyatt}, M.~C., {Bailey}, V., {Bryden}, G., {Danchi}, W.~C., {Defr{\`e}re}, D., {Haniff}, C., {Hinz}, P.~M., {Lebreton}, J., {Mennesson}, B., {Millan-Gabet}, R., {Morales}, F., {Pani{\'c}}, O., {Rieke}, G.~H., {Roberge}, A., {Serabyn}, E., {Shannon}, A., {Skemer}, A.~J., {Stapelfeldt}, K.~R., {Su}, K. Y.~L., and {Weinberger}, A.~J., ``{Exo-zodi Modeling for the Large Binocular Telescope Interferometer},'' {\em \apjs}~{\bf 216},  23 (Feb. 2015).

\end{thebibliography}
\bibliographystyle{spiebib} 

\end{document}